\documentclass[reprint,amsmath,amssymb,aps,pra]{revtex4-1}
\usepackage[utf8]{inputenc}
\usepackage{graphicx}
\usepackage{dcolumn}
\usepackage{mathrsfs,amsmath}
\usepackage{bm}
\def\ph2{{\it p}-H$_2$}
\def\Am3{\AA$^{-3}$}

\begin{document}
\title{Dipolar bosons in one dimension: the case of longitudinal dipole alignment}
\author{Youssef Kora and Massimo Boninsegni}
\affiliation{Department of Physics, University of Alberta, Edmonton, Alberta, T6G 2E1, Canada}
\date{\today}

\begin{abstract}
    We study by quantum Monte Carlo simulations the low-temperature phase diagram of dipolar bosons confined to one dimension, with  dipole moments aligned along the direction of particle motion. A hard core  repulsive potential of varying range ($\sigma$) is added to the dipolar interactio n, in order to ensure stability of the system against collapse. In the $\sigma\to 0$ limit the physics of the system is dominated by the potential energy and the ground state is quasi-crystalline; as $\sigma$ is increased the attractive part of the interaction weakens and the equilibrium phase evolves from quasi-crystalline to a non-superfluid liquid. At a critical value $\sigma_c$, the kinetic energy becomes dominant and the system undergoes a quantum phase transition from a self-bound liquid to a gas. In the gaseous phase with  $\sigma\to\sigma_c$, at low density  attractive interactions bring the system into a ``weak" superfluid regime. However,  gas-liquid coexistence also occurs, as a result of which the topologically protected superfluid regime is not approached.
\end{abstract}
\maketitle

\section{Introduction}
Dipolar gases have been an active research area in both experimental and theoretical circles, since the experimental achievement of Bose-Einstein Condensation of  atomic systems with large magnetic moments \cite{griesmeier,lu,aikawa,depaz,ni,yan,takekoshi,park,balewski}. More recently, such systems have garnered a lot of interest as possibly promising candidates for the observation of the elusive supersolid phase (for a review, see, for instance, Ref. \onlinecite{rmp}). \\ \indent
Dipolar Bose systems were suggested to underlie a possible supersolid phase in the (quasi)-2D, purely repulsive limit, i.e., with dipole moments all aligned in the direction perpendicular to the plane \cite{spivak}. 
First principle calculations have
however ruled out such an intriguing scenario, at least for practical purposes \cite{bm14}. Indeed, recent work has shown that no supersolid phase arises in 2D even if dipole moments are allowed to be tilted with respect to the perpendicular axis, which leads to the formation of striped crystals \cite{cb19}.
On the other hand, in 3D there is experimental evidence
\cite{ferlainolonglived, pfautransient,tanzi}, backed by theoretical studies \cite{cinti17,patterned,ancilotto,pohl}, of a filament supersolid phase of an assembly of bosons with aligned  dipole moments.
\\ \indent
However, systems of dipolar bosons in reduced dimensions retain significant fundamental interest, at least from a theoretical perspective, and have been extensively studied in recent times \cite{bohn,santos,pawlowski}. Of particular interest is the case in 1D, as the physics of one-dimensional many-body systems has been the subject of intense theoretical investigation for decades. A number of exact solutions and/or rigorous physical statements have been obtained \cite{mbp}, and there exists a well-established, universal theoretical framework that describes 1D systems, known as Luttinger Liquid Theory (LLT).
Considerable effort has been devoted to the realization in the laboratory of systems that may approach the 1D limit, in order to test the most important predictions of the existing theory.
\\ \indent 
Experimentally, the quasi-1D limit can be probed in different ways and/or physical settings. For example, mass flux in solid $^4$He \cite{solidhe4,solidhe42} is speculated to be essentially one-dimensional in nature, and scenarios have been proposed to the effect that a 3D supersolid phase of $^4$He may arise in a network of interconnected superfluid dislocations \cite{schev}. Alternatively,  one can adsorb
gases made of small atoms or molecules, such as helium,
inside carbon nanotubes \cite{hallock,stan}, or in porous glasses such as vycor, in which particle motion in confined to within $\sim$ 1 \AA\ in two directions \cite{beamish}. \\ \indent
This has motivated theoretical studies of hard core fluids such as $^4$He \cite {kro,bm00} and parahydrogen (p-H2) \cite{boninsegni13} in strictly 1D, as
well as inside a single nanotube \cite{he4nanopores,glyde1d,hnanopores}, or in the interstitial channel of a bundle of nanotubes \cite{crespi}.
\\ \indent
More recently, experimental advances in cold atom physics appear now to enable not only systematic, controllable confinement of particles, but also the tuning of the inter-particle interactions, e.g., through the Feshbach resonance \cite{feshbach}. This paves the way to the experimental validation of the existing theory, to an unprecedented degree of accuracy.
\\ \indent
One-dimensional systems of dipoles are of interest because the interaction, while not strictly long-ranged, has a much greater spatial extent than most conventional (i.e., atomic or molecular) interactions, and/or interactions for which analytical results are known. Moreover, it is anisotropic, which in principle can lead to different physical behavior on aligning dipole moments in different directions.
\\ \indent
\begin{figure}[h]
\centering
\includegraphics[width=0.47\textwidth]{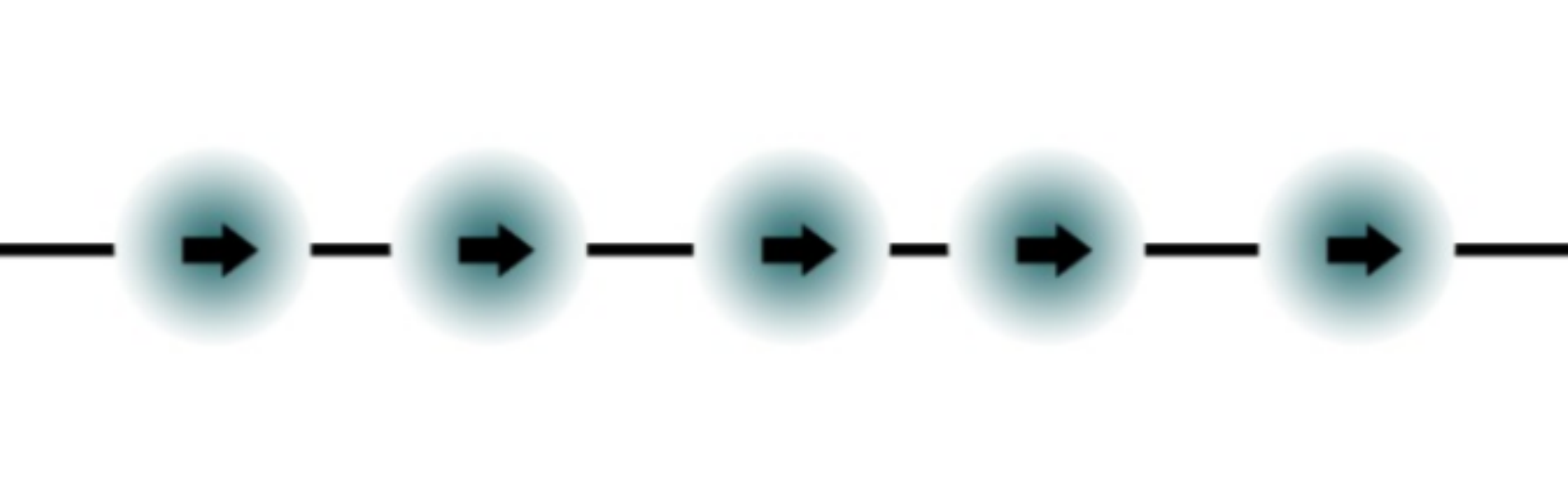} 
\caption{{\em Color online}. A system of bosons confined to move in one dimension, with dipole moments aligned parallel to the direction of motion.}
\label{dipoles}
\end{figure}
Dipolar bosons in 1D have been studied in previous works \cite{santos2,citro,reimann,roscilde,guan}, typically  in the the case of dipole moments aligned perpendicularly to the direction of particle motion, rendering  the  dipolar interaction purely repulsive, i.e., with no many-body bound state. In this paper, instead, we align the dipole moments along the direction of motion (see Fig. \ref{dipoles}), which makes the dipolar interaction purely attractive.
In order to prevent the system from collapsing, we add a hard-sphere-like repulsion of variable range $\sigma$. The presence of attractive interactions qualitatively alters the physical behavior of the system, with respect to the case studied so far. By tuning the range of the repulsive interaction, we are able to explore the different physical regimes and phases accessible to the system. 
\\ \indent
We carried out a systematic investigation of the ground state phase diagram of the system as a function of particle  density and $\sigma$, by means of computer simulations.
The main results of our study are the following: {\it a}) the system is self-bound in the $\sigma\to 0$ limit, in which the two-body interaction features a deep attractive well; the character of the many-body ground state evolves from quasi-crystalline to a non-superfluid liquid as $\sigma$ is increased, and for $\sigma > \sigma_c$, a gas-liquid quantum phase transition occurs, as the system becomes unbound {\it b}) in the gas phase, near the critical value $\sigma_c$, attractive interactions bring the system at low density into the regime known as ``weak superfluid", i.e., unstable against infinitesimal perturbations (e.g., disorder or commensurate potentials). Interestingly, the topologically protected superfluid regime cannot be approached, as the system breaks down into coexisting gas and liquid phases in the dilute limit, despite being unbound. This behavior is reminiscent of that predicted for quasi-2D $^3$He films adsorbed on weakly attractive substrates.
\\ \indent
The remainder of this paper is organized as follows: in section \ref{mo} we describe the model of the system, and briefly summarize the universal theoretical framework that describes systems in 1D; in Sec. \ref{me} we  describe our methodology; we present and discuss our results in Sec. \ref{res} and finally outline our conclusions in Sec. \ref{conc}.
\\ \indent

\section{Model}\label{mo}
We model the system as an ensemble of $N$ identical particles of mass $m$ confined to the $x$-axis. The particles have spin zero, i.e., they obey Bose statistics, and a magnetic moment $d$ pointing in the positive $x$-direction. The system is enclosed in a box of length $L$ with periodic boundary conditions. The density of the system is $\rho=N/L$. We take the characteristic length of the dipolar interaction, $a \equiv md^2/\hbar^2$ as our unit of length, and $\epsilon \equiv \hbar^2/(ma^2)$, as that of energy. In these units, the dimensionless quantum-mechanical many-body Hamiltonian reads as follows:
\begin{eqnarray}\label{u}
\hat H = - \frac{1}{2} \sum_{i}\frac{\partial^2}{\partial x^2_i}+\sum_{i<j}U(x_i,x_j)
\end{eqnarray}
where the first (second) sum runs over all particles (pairs of particles). The pair potential comprises two parts, 
\begin{equation}
U(x,x^\prime) = U_{d}(x,x^\prime) + U_{sr}(x,x^\prime).
\end{equation} 
$U_{d}$ is the classical dipolar interaction, which, for particles confined to the x-axis with their dipole moments pointing along the same axis, reads
\begin{equation}\label{Ud}
U_d(x)=-\frac{2}{|x|^3},
\end{equation}
i.e., unlike the case in which dipoles are aligned in the direction perpendicular to the line of particle motion, it is purely attractive and 
would lead to the collapse of the system, if a short-range, repulsive part were not included in the interaction. The  physical  origin  of  such  a  repulsive  term  can be different, depending on the physical system. Any atomic
or  molecular  interaction must feature  a  hard  core  repulsion
at short distance arising from Pauli exclusion principle,
which prevents electronic clouds of different atoms from
overlapping spatially.  In that case, the effective hard core diameter is
$\sim 1$ \AA, i.e., much smaller than
the typical value of the characteristic dipolar length in the majority of current experiments (see, for instance, Ref. \onlinecite{patterned}) with cold
dipolar atoms or molecules.  Significantly greater ranges
could be achieved, e.g., by means of the Feshbach resonance \cite{feshbach}.
\\ \indent 
As in recent simulations works \cite{cinti17,patterned}, $U_{sr}$ is modeled here through the repulsive part of the standard Lennard-Jones potential, i.e.,
\begin{equation}\label{Usr}
U_{sr}(x)=(\sigma/x)^{12}
\end{equation}
The effective diameter  $\sigma$ can be directly related to the scattering length $a_s$ in 3D (see, for instance, Ref. \onlinecite{flugge}).
It is worth clarifying that the actual form of $U_{sr}$ is not expected to be important; it can be regarded as a hard wall, its role being exclusively that of preventing system collapse. The physics of interest here takes place at average interparticle separations that are significantly greater than $\sigma$, rendering the contribution of $U_{sr}$ usually relatively small at the densities of interest. For alternative ways to treat the short-range interaction, see, for instance, Ref. \cite{reimann,santos2,negretti}.
\\ \indent 
The presence of both a repulsive and an attractive term, with very different dependencies on the interparticle distance, make it possible for the system to be self-bound. In the $\sigma\to 0$ limit, $U$ features a deep attractive well, $\sim -2/\sigma^3$, as a result of which the ground state of the system is a nearly classical crystal. On the other hand, as $\sigma$ grows the attraction is progressively weakened; one expects the ground state of the self-bound system to become liquid-like, and that a liquid-gas quantum phase transition should occur for $\sigma>\sigma_c$, $\sigma_c$ being the upper limit for the existence of a self-bound state.
\\ \indent
We study the ground phase diagram of (\ref{u}) by means of computer simulations, and interpret our results in terms of the LLT, i.e., the comprehensive theoretical apparatus that describes the physics of quantum many-body systems in one dimension \cite{haldane}.
The essence of the LLT is embodied in an effective quadratic Hamiltonian, expressed in terms of two bosonic fields $\theta(x)$ and $\phi(x)$, related to density and phase oscillations respectively (see, for instance, Ref. \onlinecite{cazalilla}). In terms of these fields, to leading order, the Hamiltonian reads 
\begin{equation}\label{lutt}
H = \frac{c}{2\pi} \int_0^L dx\left[\frac{1}{K}(\partial_x\phi)^2 +K(\partial_x\theta)^2  \right]
\end{equation}
where $c$ is the speed of sound of the linearly dispersed low energy excitations, and $K$ is the universal Luttinger parameter which  characterizes the relative strength of density and phase oscillations. 
\\ \indent
In one dimension, quantum fluctuations are strong enough to destroy long-range order. However, correlation functions decay algebraically, allowing for the possibility of quasi-long-range order, depending on the value of the decay exponent, which is $K$ ($1/K$) in the case of phase (density) correlations; broadly speaking, in the $K \to 0$ ($K \to \infty$) limit the system possesses quasi-superfluid (quasi-crystalline) order \cite{warning}. We come back to this point below, with a more precise classification.
\section{Methodology}\label{me}
As mentioned above, we carry out computer simulations of the system described in section \ref{mo} using the continuous-space worm algorithm \cite{worm,worm2}. We shall not review the details of this method, referring instead the reader to the original references. We utilized a canonical variant of the algorithm in which the total number of particles $N$ is held constant, in order to simulate the system at fixed density \cite{mezz1,mezz2}. Although this is a finite temperature ($T$) technique, we perform simulations at sufficiently low $T$ so that the results can be regarded as essentially ground state (we come back to this below). Experience accumulated over the past two decades shows that finite temperature techniques are a reliable options to study the ground state of Bose systems, as they are unaffected by serious limitations plaguing ground state methods \cite{hinde,population,psb}.
\\ \indent
Although the computational methodology adopted here allows for the calculation of off-diagonal correlations, the results shown in Sec. \ref{res} all pertain to diagonal correlations; therefore, since exchanges of indistinguishable particles are strictly forbidden in 1D by the hard core of the interaction, they can be obtained by means of conventional path integral Monte Carlo as well (see, for instance, Ref. \onlinecite{jltp}). 
\\ \indent 
In the ground state, the physics of the system depends exclusively on the value of $\sigma$ and on the density $\rho$, or, equivalently, the average interparticle separation $\lambda\equiv \rho^{-1}$.
We investigate the ground state of the system as a function of $\sigma$; that is, for a given value of $\sigma$, we compute the equation of state,  determining the equilibrium density of the system, and at that density compute relevant correlation functions in real and momentum space. We performed simulations for values of $\sigma$ in the [0.10,2] range; our typical system sizes range from $N=25$ to $N=400$. 
Details of the simulation are standard; we made use of the fourth-order approximation for the high-temperature density matrix, ll of the results quoted here are extrapolated to the limit of time step $\tau\to 0$.
\\ \indent
As mentioned above, while true long-range order cannot exist in 1D, quasi-order can manifest itself in the form of an algebraic decay of the correlation functions, governed by the exponent $K$. By studying the evolution of its value as a function of $\sigma$ (i.e., of the equilibrium density) we characterize the kind of (quasi) order that the system displays. Two methods were primarily utilized to extract the Luttinger parameter $K$ at the various physical conditions. The first is through the static structure factor $S(q)$, which quantifies the strength of density fluctuations with momentum $q$. In the units adopted here, the Luttinger parameter $K=c\lambda/\pi$, where $\lambda=1/\rho$ is the interparticle distance and $c$ is the speed of sound, accessible from the long-wave behavior of the static structure factor through relation
\begin{equation}\label{slope}
\frac{1}{2c} = \lim_{q\to0} \frac{S(q)}{q}
\end{equation}
\\ \indent
The second method to calculate $K$ is through the equation of state $e(\rho)$, i.e., the energy per particle as a function of the density at $T=0$, from which one can obtain the compressibility $\kappa=\rho^{-1} \ \partial \rho/\partial P$, where $P$ is the pressure. $\kappa$ is related to $K$ through the relation $K = (\pi^2\ \lambda_\circ^3\ \kappa)^{-1/2}$, where $\lambda_\circ$ is the interparticle distance at the equilibrium density \cite{cazalilla}. The agreement of the values of $K$ computed through the two methods serves as a self-consistency check.

\section{Results}\label{res}

\begin{figure}[h]
\centering
\includegraphics[width=0.47\textwidth]{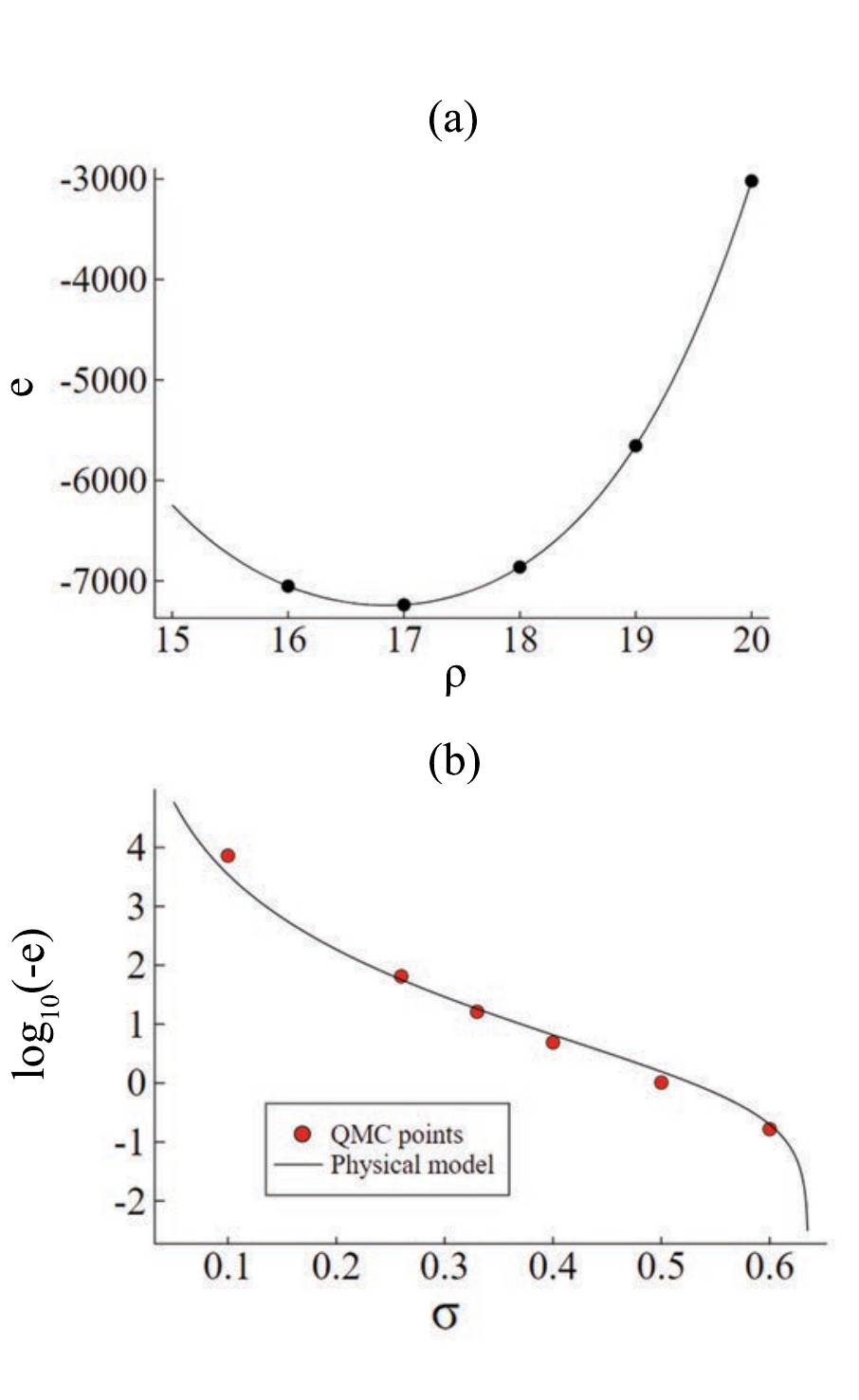} 
\caption{{\em Color online}. (a) Energy per particle $e(\rho)$ at $\sigma=0.10$ as a function of density in the $T\to0$ limit. Statistical errors are smaller than the size of the symbols. Solid line is a quartic fit to the data. (b) Logarithm of the negative energy per particle in the $T\to0$ limit, as a function of $\sigma$. Circles are the results of the simulations, solid line is a fit based on Eq. \ref{ex} (see text).}
\label{energy}
\end{figure}

We compute the equation of state $e(\rho)$, where $e$ is the energy per particle, in the $T=0$ limit, as a function of $\sigma$. As ours is a finite temperature method, extrapolation of the results obtained at low $T$ is required; in practice, the shape of the curve $e(\rho)$ is found not to change significantly once the temperature is of order $\sim 0.1$ of the average kinetic energy per particle.
\\ \indent
An example is shown in Fig. \ref{energy}a for $\sigma=0.10$. For this value of $\sigma$, the interaction potential possesses a deep attractive well, and as a result $e(\rho)$ displays a clear minimum at $\rho=17$, which corresponds to the equilibrium density. Simulations at density lower than the equilibrium one can be carried out down to the spinodal density, i.e., that at which the speed of sound vanishes and below which the uniform system becomes unstable against the formation of ``puddles'' of fluid. The curvature of $e(\rho)$ also provides a method of computing the value of the Luttinger parameter, as explained in Sec. \ref{me}. 
\\ \indent
As the value of $\sigma$ is increased, the magnitude of the binding energy decreases, until it hits zero at $\sigma=\sigma_c$, whereupon the system becomes unbounded. In order to obtain a quantitative estimate for $\sigma_c$, we fit our computed $e(\rho)$ for the different values of $\sigma$, with the following simple expression, based on the crude approximation for the pair correlation function $g(x)=\Theta(x-\lambda/2)$ ($\Theta$ being the Heaviside's function):
\begin{equation}\label{ex}
e(\lambda) = \frac{C_1}{\lambda^2} - \frac{C_2}{\lambda^3} + 0.045\left(\frac{2\sigma}{\lambda}\right)^{12},
\end{equation}
where $C_1$ and $C_2$ are fitting parameters.  The results are shown in Fig. \ref{energy}b. Beyond $\sigma=\sigma_c$, the first (kinetic energy) term in Eq. \ref{ex} overtakes the second  (attractive part of the potential energy) in magnitude, the system is no longer self-bound and a liquid-gas quantum phase transition occurs. We estimate $\sigma_c = 0.65 \pm 0.02$, based on the values of the fitting parameters $C_1\approx 1.25$ and $C_s\approx 0.76$.  
\begin{figure}[h]
\centering
\includegraphics[width=0.47\textwidth]{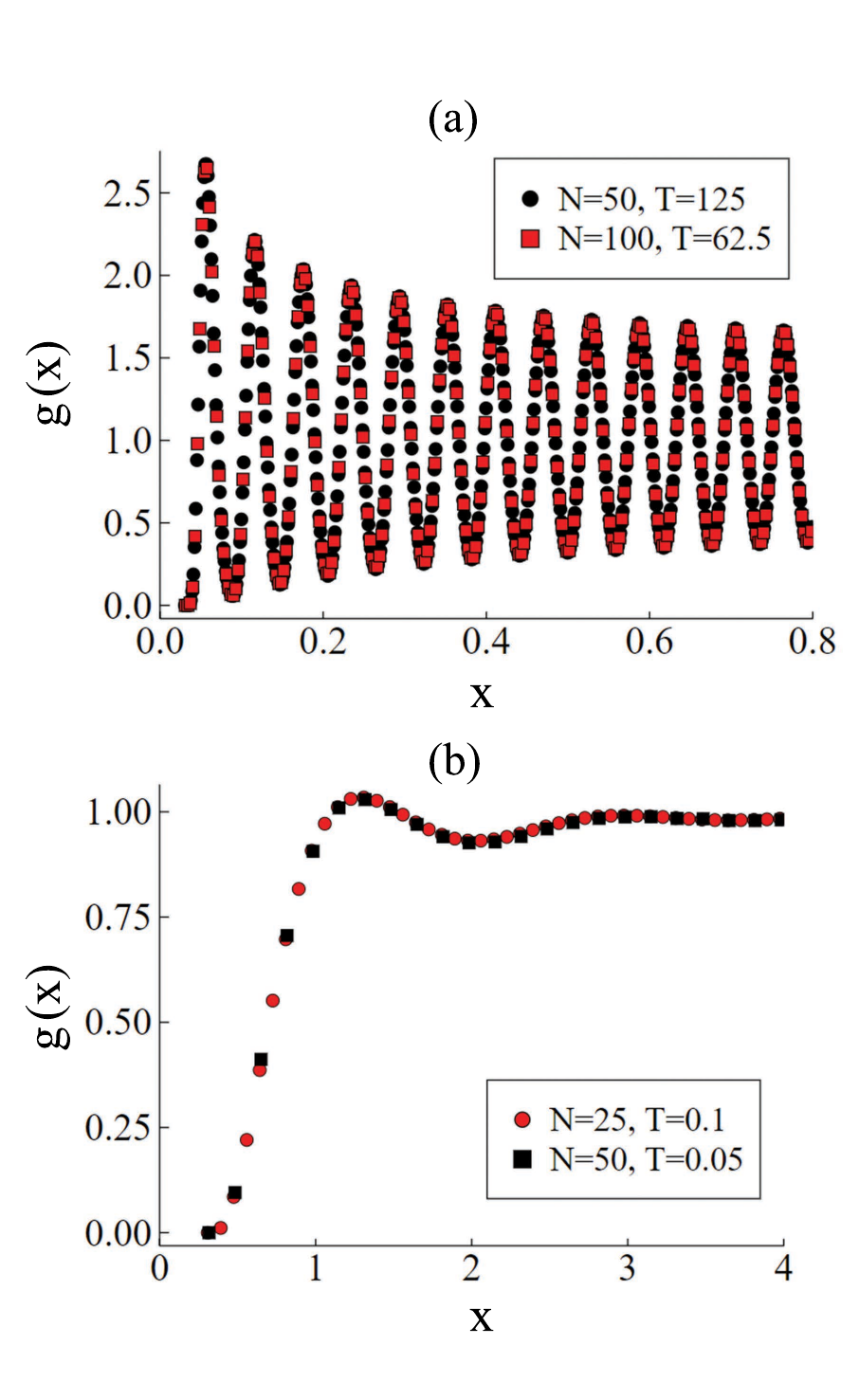} 
\caption{{\em Color online}. Ground state pair correlation functions $g(x)$ for the system at $\sigma=0.10$ (a) and $\sigma=0.60$ (b), each at two different system sizes $N$ and temperatures $T$, at the computed equilibrium densities, namely $\rho =17\ (0.6)$ for $\sigma=0.10\ (0.60)$ Statistical errors are smaller than the size of the symbols.}
\label{gx}
\end{figure}

The evolution of the structure of the ground state as the hard core diameter $\sigma$ is varied can be illustrated by means of the pair correlation function $g(x)$. At low $\sigma$, the depth of the attractive well favors a quasi-crystalline orderly arrangement of particles, consistent with a value of the Luttinger parameter $K > 2$. Fig. \ref{gx}a shows the result for  $g(x)$ at $\sigma=0.10$, at the equilibrium density $\rho=0.17$. On the other hand, as $\sigma$ grows and the binding energy tends to zero, the system acquires a more liquid-like character, behaving essentially like a hard sphere fluid. This is clear in Fig. \ref{gx}b, where $g(x)$ is displayed for $\sigma=0.60$, at the equilibrium density $\rho=0.6$. In both cases, the computed $g(x)$ features the expected scaling, i.e., results at a given density only depend on the product $NT$, in the low-$T$ limit.

\begin{figure}[h]
\centering
\includegraphics[width=0.47\textwidth]{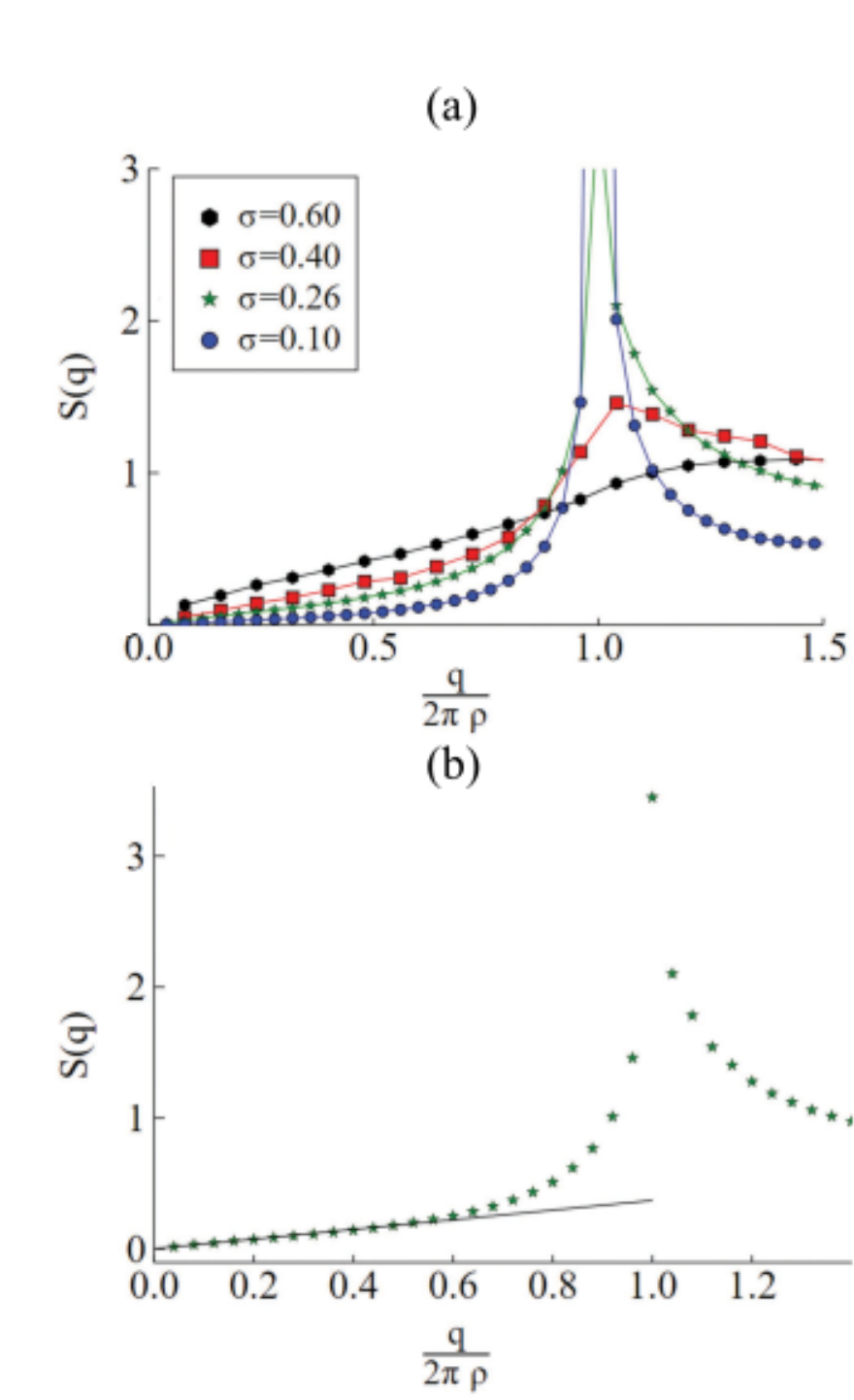} 
\caption{{\em Color online}. (a) The static structure factor at the equilibrium density for various values of $\sigma$, in the low temperature limit. Statistical errors are smaller than the size of the symbols. (b) Example of extraction of the Luttinger parameter $K$ based on Eq. \ref{slope}, for $\sigma=0.26$.}
\label{sq}
\end{figure}
A more quantitative characterization of the physics of the ground state of the system is achieved through the determination of the Luttinger parameter $K$. We discuss our results for the Luttinger parameter for the two cases $\sigma<\sigma_c$ (i.e., where the system is self-bound) and $\sigma>\sigma_c$. As explained above, the most direct way of obtaining $K$ from the simulation data makes use of the computed static structure factor $S(q)$, through Eq. \ref{slope}.
Results for the static structure factor at the equilibrium density for different values of $\sigma < \sigma_c$, are shown in Fig. \ref{sq}. $S(q)$ is computed directly and/or through  the Fourier transform of $g(x)$. 
Fig. \ref{sq}b illustrates an example of the calculation of $K$ through $S(q)$, in this case at $\sigma=0.26$.
\begin{figure}[h]
\centering
\includegraphics[width=0.47\textwidth]{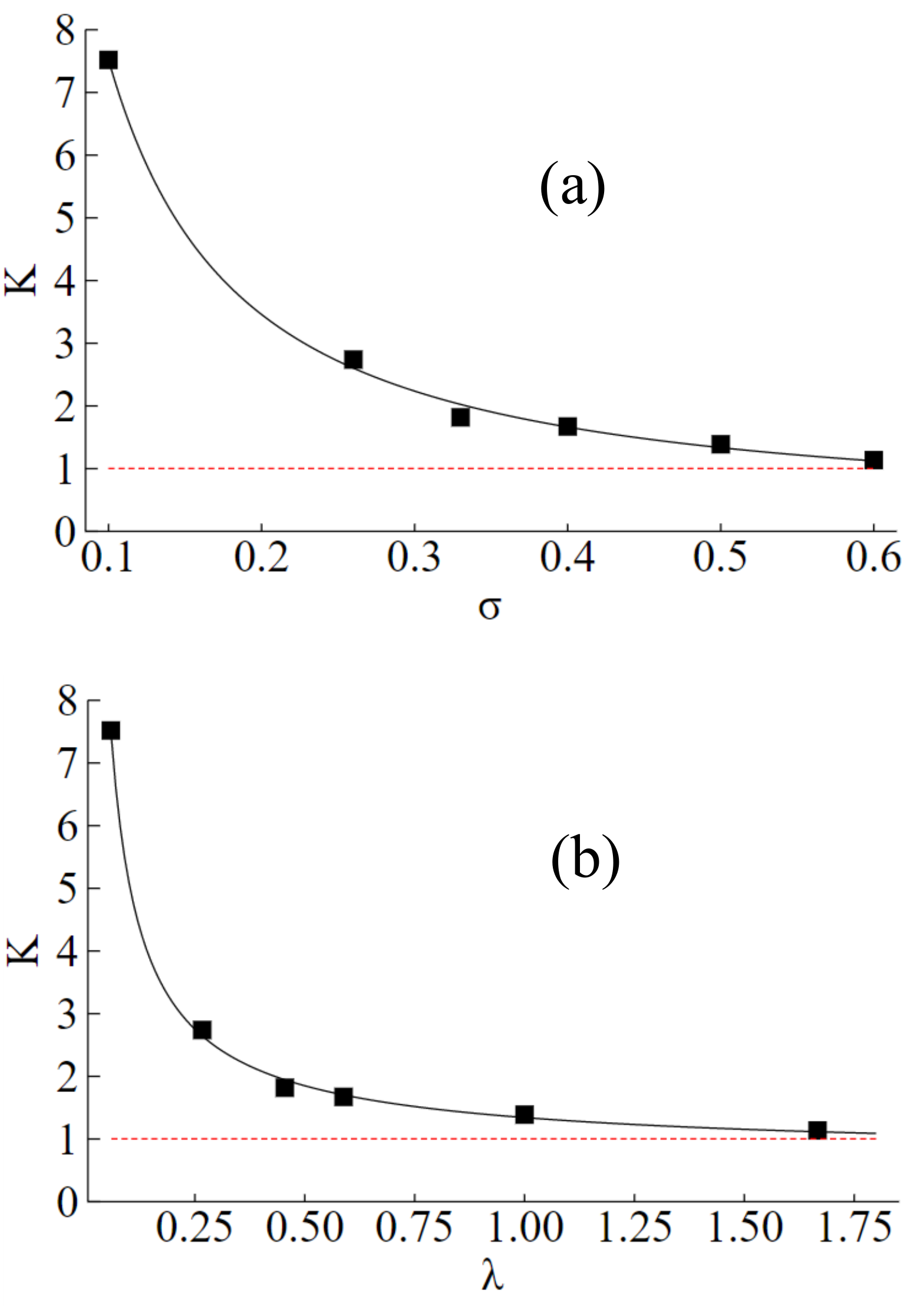}
\caption{{\em Color online}. The Luttinger parameter of the system as a function of $\sigma$ (a) and the equilibrium density $\rho$ (b). The red dashed line is $K=1$, the Tonks-Girardeau limit. Statistical errors are smaller than size of the symbols. Solid lines are guides to the eyes.}
\label{k}
\end{figure} 
\\ \indent
A strongly oscillatory behaviour of the $g(x)$, e.g., as shown in Fig. \ref{gx}a, is reflected by the appearance of divergent peaks in $S(q)$. Correspondingly, the value of $K$ is above 2, consistently with the presence of quasi-crystalline order.  As $\sigma$ grows, $K$ decreases, and becomes less than 2 at $\sigma \gtrsim 0.35$.
\\ \indent
Our results for the Luttinger parameter below $\sigma_c$ are shown in Fig. \ref{k}. As expected, $K$ at the equilibrium density is a monotonically decreasing function of $\sigma$, but was always observed to remain above 1 in the range of $\sigma$ within which the system is self-bound. It approaches unity from above as $\sigma\to\sigma_c$, above which the behavior of the system is dominated by repulsive interactions (the  Tonks-Girardeau regime  \cite{cazalilla}). 
For $0.35 \lesssim\sigma\lesssim\sigma_c$, it is $2 > K > 1$, i.e., no evidence was found of topologically protected superfluid phases in the range in which the system exists as a self-bound liquid. This overall physical behavior is qualitatively distinct from that of both 1D $^4$He, which is a quasi-superfluid at equilibrium \cite{bertaina}, as well as parahydrogen, which is quasi-crystalline \cite{boninsegni13}. 
\begin{figure}[h]
\centering
\includegraphics[width=0.47\textwidth]{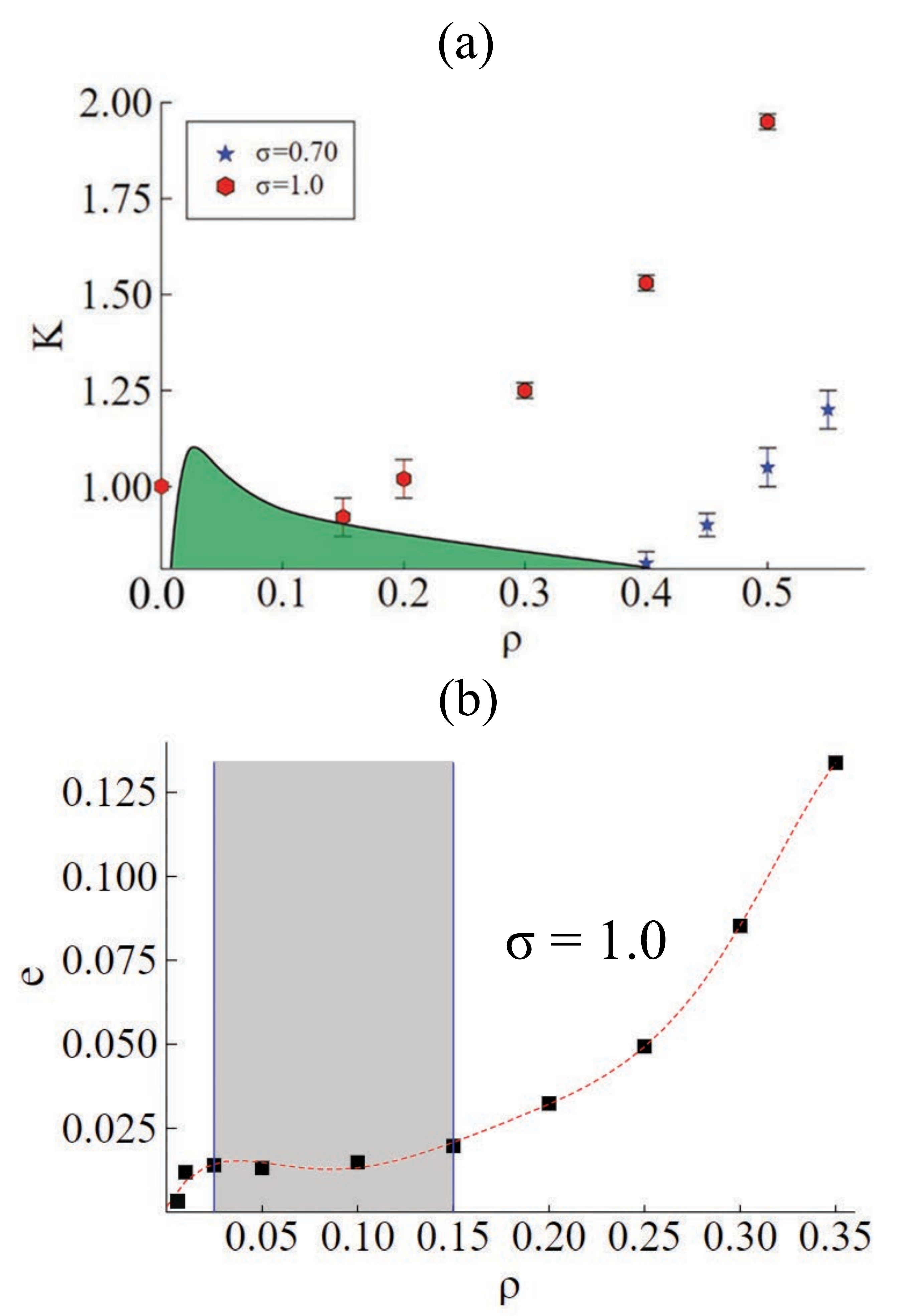}
\caption{{\em Color online}. (a) The Luttinger parameter of the system as a function of $\rho$ for two values of $\sigma>\sigma_c$. The shaded area corresponds to the (speculated) region of the phase diagram where phase coexistence occurs. (b) Energy per particle $e(\rho)$ at $\sigma=1$ as a function of the density. The shaded area corresponds to the range of densities where the system is found to feature phase coexistence. Statistical errors are smaller than size of the symbols.}
\label{ps}
\end{figure} 
\\ \indent
 Our results for the Luttinger parameter for $\sigma \gtrsim\sigma_c$ are shown in Fig. \ref{ps} for two values of $\sigma$, namely $0.7$ and $1$.  The first value of $\sigma$ is in the immediate vicinity of $\sigma_c$. While the system is no longer self-bound, the attractive part of the interaction drives $K$ below 1, as in the case of  1D $^3$He \cite{boronat}. The lowest computed value of $K$ is $\approx 0.8$ for $\rho=0.4$; at lower density, the system is observed to break down into two coexisting phases, a low-density gas and a liquid of density $\rho=0.4$. The same behavior is observed for $\sigma=1$, for which $K$ reaches a minimum value of $\sim 0.9$ for $\rho \approx 0.15$.
\\ \indent
 The coexistence of a low-density gas and a liquid phase in a system that is not self-bound was already reported in quasi-2D $^3$He films on weakly attractive substrates \cite{ruggeri}; it is reflected in the equation of state, as  shown in Fig. \ref{ps}b for $\sigma=1$. The energy per particle is a nearly ``flat" function of the density \cite{finite} in the range $0.025<\rho<0.15$. The width of this region of phase coexistence is found to diminish as the value of $\sigma$ is increased, and for $\sigma$ as high as 1.5 no coexistence is observed, but K remains above 1. This is illustrated in Fig. \ref{ps}a, which features a roughly sketched shaded area corresponding to the speculated shape of the region of phase coexistence. It should be emphasized that we did not carry out a quantitative investigation of the boundaries of the region of phase coexistence, i.e., we are unable to say at what low density the system returns to a homogeneous, low-density gas phase, for a given value of $\sigma$. The physics of such a dilute phase is expected to be amenable to a description in terms of a Tonks-Girardeau gas.
\section{Conclusions}\label{conc}
We carried out quantum Monte Carlo simulations of a system of spin zero dipolar particles in one dimension, with their dipole moments aligned along the direction of motion. The interaction also includes a hard core, repulsive term (with a variable range $\sigma$), in order to ensure thermodynamic stability of the system against collapse. The phase diagram of the system is found to display considerably more richness with respect to the case previously investigated in previous works, with the dipoles aligned perpendicular to the direction of particle motion. In the latter case, the interaction is purely repulsive and the system has no self-bound regime, excluding the possibility of a quantum phase transition. Moreover, as reported in Ref. 37, the system with perpendicularly aligned dipoles has a value of K always above 1. Our system, on the other hand,  features quasi-crystalline order at very low values of $\sigma$, and evolves into a non-superfluid liquid as $\sigma$ grows. Beyond a critical value $\sigma_c$, inter-particle attraction is no longer sufficient to keep the system self-bound. Slightly above $\sigma_c$ and on lowering the density, superfluidity may be achieved, albeit topologically unprotected and unstable against disorder. Further lowering the density results in gas-liquid phase coexistence.  It is worth emphasizing that the investigation carried out here is not merely of academic interest, as the value of $\sigma$ is experimentally tunable, and so our predictions are in principle testable in the laboratory. Obviously, important issues have to be taken into account as the 1D limit is approached. e.g., the importance of transverse modes \cite{citro3}. 

\section*{acknowledgments}
This work was supported by the Natural Sciences and Engineering Research Council of Canada. Computing support of Compute Canada is gratefully acknowledged. 


\begin{thebibliography}{99}
\bibitem{warning}
It should be mentioned that, while the convention adopted here is the same as in many other works, elsewhere in the literature $K$ is defined as the reciprocal of the parameter defined in this work.
\bibitem{finite}
Although this curve is computed in this work at finite temperature, it  retains its shape in the low temperature limit.
\end{thebibliography}
\end{document}